\documentclass{article}
\usepackage[utf8]{inputenc}
\usepackage[lmargin=0.65in,rmargin=0.65in,tmargin=0.9in,bmargin=0.8in]{geometry}
\usepackage{hyperref,lmodern,textcomp,graphicx,float,caption,fancyhdr}
\usepackage[table]{xcolor}
\usepackage{color,amsmath,amssymb,mathpazo,mathtools,bbm}
\usepackage{vwcol}
\usepackage{lipsum}
\usepackage{tgpagella}
\usepackage{indentfirst}
\usepackage{bchart}
\usepackage{lettrine,multicol}
\usepackage[sorting=none,style=phys,autocite= superscript]{biblatex}
\usepackage{titlesec}

\titleformat*{\section}{\large\bfseries}
\titleformat*{\subsection}{\normalsize\bfseries}
\titleformat*{\subsubsection}{\normalsize\itshape}

\usepackage{tcolorbox}%
\definecolor{mycolor}{rgb}{0.122, 0.435, 0.698}% Rule colour
\makeatletter
\newcommand{\mybox}[1]{%
  \setbox0=\hbox{#1}%
  \setlength{\@tempdima}{\dimexpr\wd0+13pt}%
  \begin{tcolorbox}[boxrule=0pt,arc=0.pt,
      left=6pt,right=6pt,top=6pt,bottom=6pt,boxsep=1pt]
    #1
  \end{tcolorbox}
}

\fancypagestyle{firstpage}
{\fancyhead[L]{\textsc {\color {gray} {\large }}}    
\fancyhead[R]{\large \textsc{\color {gray} {Short Communication}}}}
    
\begin{document}

\thispagestyle{firstpage}
{\noindent \Large \textbf{Onset of wavenumber bandgaps via alternating Willis coupling signs}} \\ [0.5em] 

\noindent {\large {Hasan B. Al Ba'ba'a}} \\

\noindent \begin{tabular}{c >{\arraybackslash}m{6in}}
    \includegraphics[]{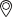} &
    \noindent {\small Department of Mechanical Engineering, Union College, Schenectady, NY 12308, USA} \\[0.25em]
    
    \includegraphics[]{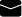}& \noindent {\href{mailto:albabaah@union.edu}{\small albabaah@union.edu}}\\
    
\end{tabular}

%%%%%%%%%%%%%%%%%%%%%%%%%%%%%%%%%%%%%%%%%%%%%%%%%%%%%%%%%%%%%%%%%%%%%%%%%%%%%%%%%%%%%%%%%%%%%%%%%%%%%%%%%%%%%%%%%%%%%%%%%%%%%%%%%%%%%%%%%%

\vspace{0.5cm}

\noindent\rule{7.2in}{0.5pt}

\mybox{\textit{Abstract} --- \small{This article introduces a methodology for inducing wavenumber bandgaps via alternating Willis coupling signs.~A non-reciprocal wave equation of Willis-type is first considered, and its wave dispersion analyses are carried out via the transfer matrix method.~By creating unit cells from two identical Willis-type elastic layers, yet with reversed Willis-coupling signs, a reciprocal band structure peculiarly emerges, although each layer exhibits non-reciprocity if considered individually.~Wavenumber bandgaps open due to such unit cell configuration, and their width and limits are analytically quantified.~Similarities between materials with reversed-sign Willis coupling and bi-layered phononic crystals are noted, followed by concluding remarks.}}

\noindent \small{\textbf{Keywords
}---
\noindent Wavenumber bandgaps, Willis coupling, non-reciprocal waves, periodic structures, dispersion diagram.}

\noindent\rule{7in}{0.5pt}

\begin{multicols}{2}

\section*{Background and motivation}
Consider a non-reciprocal wave equation, in the absence of external forces~\cite{attarzadeh2018elastic}: 
\begin{equation}
\frac{\partial^2 u}{\partial t^2} + 2v_0 \frac{\partial^2 u}{\partial x \partial t} +(v_0^2-c^2)\frac{\partial^2 u}{\partial x^2} = 0
\label{eq:modified_waveEq}
\end{equation}
governing the longitudinal displacement $u(x,t)$ of an axially moving elastic rod, quantified along the axial dimension $x$ at any instant of time $t$. The wave speed in the elastic medium $c = \sqrt[]{E/{\rho}}$ is a function of its density $\rho$ and modulus of elasticity $E$. The elastic rod is assumed to have a constant cross-sectional area $A$. Besides being of gyroscopic nature~\cite{Wickert1990MovingContinua}, Equation~(\ref{eq:modified_waveEq}) is also known as a Willis-type equation of motion, where $v_0$ symbolizes the Willis coupling coefficient~\cite{nassar2017modulated,nassar2022waves}, equivalent to the rod's speed~\cite{attarzadeh2018elastic}. Non-reciprocity in Willis-type elastic media stems from the introduced momentum bias, as evident from the mixed derivative term in Equation~(\ref{eq:modified_waveEq}).

To examine wave dispersion and understand periodic structures governed by Equation~(\ref{eq:modified_waveEq}), a transfer matrix of a unit cell of length $x$ is developed. To streamline the analysis, three non-dimensional quantities are introduced: (i) a normalized modulation speed $\nu = v_0/c$\footnote{The present study limits the value of the normalized modulation speed to the range $\nu \in [0, 1)$ to avoid possible dynamical instabilities~\cite{attarzadeh2018elastic}.}, (ii) non-dimensional length $\xi = x/\ell$, and (iii) a non-dimensional time $\tau = \omega_0 t$, with $\omega_0 = c(1-\nu^2)/\ell$ and $\ell$ being an arbitrary length.~Implementing the non-dimensional quantities modifies Equation~(\ref{eq:modified_waveEq}) to:
\begin{equation}
(1-\nu^2) \frac{\partial^2 u}{\partial \tau^2} + 2\nu \frac{\partial^2 u}{\partial \xi \partial \tau} - \frac{\partial^2 u}{\partial \xi^2} = 0
\label{eq:normalized_WaveEq1}
\end{equation}
Following the methodology in Ref.~\cite{attarzadeh2018elastic} and assuming $u(\xi,\tau) = \bar{u}(\xi) \text{e}^{-\mathbf{i} \Omega \tau}$, the following unit-cell's transfer matrix is developed, 
\begin{equation}
    \mathbf{T}(\nu) = \text{e}^{-\mathbf{i}\nu \Omega \xi} \mathbf{Y}(\nu)
    \label{eq:T_linear}
\end{equation}
where the matrix $\mathbf{Y}(\nu)$, having a determinant of one, is given by:
\begin{equation}
\mathbf{Y}(\nu) = \begin{bmatrix}
\cos \left( \Omega \xi\right) + \mathbf{i}\nu \sin\left( \Omega \xi\right) & \frac{1}{k \Omega} \sin (\Omega \xi) \\ -(1-\nu^2)k \Omega \sin (\Omega \xi) & \cos \left( \Omega \xi\right) - \mathbf{i}\nu \sin\left( \Omega \xi\right)
\end{bmatrix}
\label{eq:Y_linear}
\end{equation}
\vspace{0.4in}

In Equations~(\ref{eq:T_linear})~and~(\ref{eq:Y_linear}), $\Omega = \omega/\omega_0$ is a normalized version of the excitation frequency $\omega$, $k = EA/\ell$ is the effective longitudinal stiffness of a rod segment of length $\ell$, and $\mathbf{i} = \sqrt{-1}$ is the imaginary unit. The derived transfer matrix $\mathbf{T}(\nu)$ relates the state vectors at the terminals of the unit cell according to $\mathbf{z}(\xi) =
\mathbf{T}(\nu)
\mathbf{z}(0)$, where the state vector $\mathbf{z}(\xi)= \{\bar{u}(\xi) \ f(\xi) \}^\text{T}$ compiles the displacement $\bar{u}(\xi)$ and internal force $f(\xi) = k \bar{u}_\xi$ at a given terminal at distance $\xi$. 

Despite the introduction of a momentum bias, I next show the linearity of the dispersion relation from analyzing the eigenvalues of transfer matrix $\mathbf{T}(\nu)$ in Equation~(\ref{eq:T_linear}) with $\xi = 1$. Computing the determinant of $|\mathbf{T}(\nu)-\lambda \mathbf{I}| = 0$, where $\lambda$ is an eigenvalue and $\mathbf{I}$ is a $2 \times 2$ identity matrix, the following solutions of $\lambda$ are derived:
\begin{equation}
    \lambda = \text{e}^{-\mathbf{i}\nu \Omega} \left( \frac{\text{tr}(\mathbf{Y})}{2} \pm \mathbf{i} \sqrt{1- \left(\frac{\text{tr}(\mathbf{Y})}{2} \right)^2} \right)
\label{eq:lambda}
\end{equation}
If the eigenvalues are written as $\lambda = \text{e}^{\mathbf{i}q}$, where $q$ is a non-dimensional wavenumber, the driven-wave dispersion relation is developed from Equation~(\ref{eq:lambda}) after executing a few mathematical manipulations:
\begin{equation}
    q = q_s \pm \Omega
\label{eq:linear_disp}
\end{equation}
where $q_s = -\nu \Omega$ is a wavenumber \textit{phase shift}, reminiscent of the phase shift observed in Willis monatomic lattices recently developed by the author~\cite{al2023Brillouin}. As evident from Equation~(\ref{eq:linear_disp}), and in agreement with the results obtained in Ref.~\cite{attarzadeh2018elastic}, the dispersion relation remains linear, even with the presence of Willis coupling (or rod's motion). However, the slopes of the dispersion branches (i.e., group velocity) corresponding to forward-going and backward-going waves are unequal, indicative of non-reciprocal wave propagation, and the change in slope is governed by the normalized modulation speed $\nu$.

\begin{figure}[H]
     \centering
\includegraphics[]{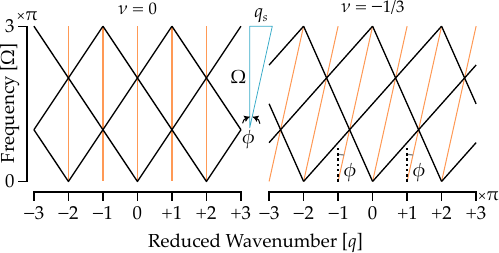}
\caption{A comparison between reciprocal ($\nu = 0$) and non-reciprocal ($\nu = -1/3$) band structures, corresponding to the wave equation in Equation~(\ref{eq:modified_waveEq}), having the dispersion relation in Equation~(\ref{eq:linear_disp}). The definition of the skew angle $\phi$ can be perceived as an angle in a triangle, constructed via the wavenumber phase shift $q_s$ and its corresponding frequency $\Omega$.~Note that the dispersion relation is repeated periodically along the wavenumber axis for better visualization of the skew angle.}
     \label{fig:linear_disp}
\end{figure}

An additional observation I note here is that the angle at which the linear dispersion skews relative to the frequency axis can be analytically quantified via a straightforward trigonometry. More specifically, a triangle is formulated via the phase shift $q_s$ and its corresponding frequency $\Omega$ as shown in Figure~\ref{fig:linear_disp}, from which the skew angle $\phi$ can be derived:
\begin{equation}
    \phi = \tan^{-1}(\nu)
\end{equation}
This skew angle $\phi$ grows non-linearly starting from $\nu = 0$, the reciprocal case, for which the triangle understandably collapses to a line. Note that, unlike the phase shift $q_s$, the angle $\phi$ is independent of the frequency $\Omega$. A comparison between the reciprocal case ($\nu = 0$) and a non-reciprocal one ($\nu = -1/3$) is shown in Figure~\ref{fig:linear_disp}.

As detailed earlier, a homogeneous elastic rod with a single Willis coupling results in non-reciprocity. One might then wonder: What is the effect of having a periodic variation in Willis coupling?~{\color{black}{Unlike elastic periodic structures with \textit{frequency} bandgaps, such as rods with variation in mechanical properties~\cite{AlBabaa2016a,hussein2007dispersive,nouh2024role}, beams with periodically mounted local resonators ~\cite{el2018discrete,liu2012wave,Nouh2014}, periodic gyroscopic systems~\cite{Nash2015TopologicalMetamaterials,Garau2018InterfacialSpinners,alsaffar2018band,attarzadeh2019non}, tensegrity chains~\cite{placidi2024variational,placidi2024bandgap,amendola2018tuning}, granular microstructures~\cite{misra2016granular,nejadsadeghi2019axially,nejadsadeghi2020role,misra2019longitudinal}, among others~\cite{Hussein2014,wang2020tunable,Jin2021pillaredPnCs}}}, this work establishes that a homogeneous elastic rod with periodic alternating signs of Willis coupling opens \textit{wavenumber} bandgaps.~The presented approach adds to typical methods for initiating wavenumber bandgaps, encompassing structural damping~\cite{hussein2010band}, temporal stiffness modulation~\cite{trainiti2019time}, supersonic spatiotemporal modulation~\cite{attarzadeh2018elastic}, and the use of generally complex spatiotemporal harmonic functions for  stiffness~\cite{Moghaddaszadeh2022Complex}.

\begin{figure}[H]
     \centering
\includegraphics[]{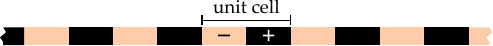}
\caption{Schematic of an infinite elastic rod with a unit cell of alternating negative/positive Willis coupling.}
     \label{fig:sch}
\end{figure}

\section*{Elastic rod with periodic Willis coupling}

Consider a unit cell of an elastic rod divided evenly into two segments as seen in Figure~\ref{fig:sch}. Each segment has equal normalized length $\xi = 1/2$, cross-sectional area, and material properties, yet reversed signs of Willis coupling.~The transfer matrix for such a unit cell can be found by multiplying the transfer matrix for each segment (i.e., Equation~(\ref{eq:T_linear})), which results in:
\begin{equation}
    \mathbf{T} = \text{e}^{-\frac{1}{2}\mathbf{i}(\nu_+ - \nu_-) \Omega}  \mathbf{Y}(\nu_+) \mathbf{Y}(-\nu_-)
\label{eq:T_matrix_UC}
\end{equation}
where $\nu_-$ and $\nu_+$ denote the Willis couplings of the negative and positive segment, respectively. As inferred from Equation~(\ref{eq:T_matrix_UC}), setting $\nu_+ = \nu_- = \nu$ reduces the phase parameter $\text{e}^{-\frac{1}{2}\mathbf{i}(\nu_+ - \nu_-) \Omega}$ to one as a consequence of equal magnitude and opposite signs of Willis couplings.~Although each segment is \textit{non-reciprocal} on its own, the lack of a phase parameter renders a zero skew angle, pertaining to a \textit{reciprocal} dispersion relation. Following identical procedure to the development of Equation~(\ref{eq:linear_disp}), the driven-wave dispersion relation of an elastic rod with alternating signs of Willis coupling (and equal magnitude) is:
\begin{equation}
    q = \pm \cos^{-1} \left(\nu^2 + \left(1 - \nu^2 \right)\cos(\Omega) \right)
\label{eq:simplified_disp_rel}
\end{equation}
As expected, no phase shift in the wavenumber is observed from Equation~(\ref{eq:simplified_disp_rel}). Interestingly, the proposed rod design yields \textit{wavenumber} bandgaps when $\nu \neq 0$. At $\Omega = (2p-1)\pi$, where $p\in \mathbb{N}$ and $\mathbb{N}$ is the set of natural numbers, wavenumber bandgaps initiate around odd integer multiples of $q = \pm\pi$, and the gaps' width is proportional to $\nu$'s magnitude.~To observe the wavenumber bandgaps, Equation~(\ref{eq:simplified_disp_rel}) solutions are repeated periodically along the wavenumber axis every $2\pi$, necessarily generating wavenumber bandgaps of \textit{identical} width, with their lower and upper limits, respectively, being:
\begin{subequations}
\begin{align}
    q_- & = \pm \left[ 2(n-1)\pi + \cos^{-1}(2\nu^2-1) \right] \\
    q_+ & = \pm \left[ 2n\pi - \cos^{-1}(2\nu^2-1) \right]
\end{align}
\label{eq:qbG-limits}
\end{subequations}
Note that $n \in \mathbbm{N}$ denotes the order of the wavenumber bandgap. Using Equation~(\ref{eq:qbG-limits}), a formula for the identical wavenumber-bandgaps' width is deduced:
\begin{equation}
   \Delta q = 2\left( \pi-\cos^{-1}\left(2 \nu^2 - 1 \right) \right)
\end{equation}

\begin{figure*}[]
     \centering
\includegraphics[]{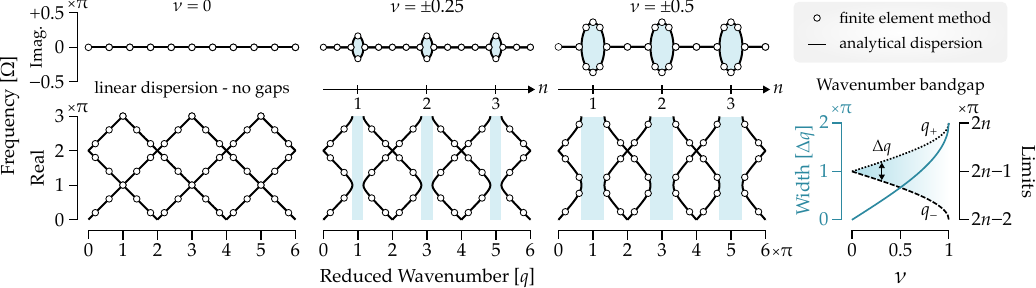}
\caption{(\textit{left}) Dispersion diagrams for the uniform rod with linear dispersion and no gaps ($\nu = 0$) and a rod with Willis coupling of alternating signs and $\nu = \pm0.25, \pm 0.5$. As can be seen, the wavenumber bandgaps open (shaded areas) whenever $\nu \neq 0$ and grows in size with increasing its magnitude, regardless of the sign. All orders of wavenumber bandgaps (denoted by $n$) have identical width. Frequencies are complex within said gaps and have larger imaginary values with higher $|\nu|$. Finite element simulations (depicted as circles) provide further validation to the analytical results.~(\textit{right}) Wavenumber-bandgap width $\Delta q$ and limits $q_\pm$ as a function of $\nu$, corroborating the behavior of bandgap growth in the dispersion diagrams.}
     \label{fig:qBG}
\end{figure*}

Figure~\ref{fig:qBG} shows band structure examples of a material with no Willis coupling (as the baseline) and with reversed-sign Willis couplings with $\nu = \pm 0.25 \text{ and} \pm 0.5$, showing the emergence of wavenumber bandgaps as predicted by Equation~(\ref{eq:qbG-limits}). As can be seen, flipping the sign of $\nu$ does not affect the band structure as it results in a different choice of the unit cell.~Further, complex frequencies emerge in wavenumber bandgap~\cite{Moghaddaszadeh2022Complex}, and the magnitude of the imaginary part grows larger with higher magnitudes of $\nu$.~The dispersion analyses are verified via the finite element method (depicted as circles in the figure), which agrees excellently with the analytical predictions. For reference, the wavenumber bandgap limits $q_\pm$ and corresponding width are also shown for a swept range of $\nu$. As $\nu$ approaches the limiting case of 1, the dispersion curves become vertical lines and the wavenumber-bandgap width peaks at $\Delta q = 2\pi$.

%  using an in-house MATLAB code and quadratic rod elements

\begin{table*}[]
\small
\caption{Comparison between the expressions for the limits and width of wavenumber bandgaps for materials with alternating signs of Willis couplings (Figure~\ref{fig:sch}) and those of frequency bandgaps of a bi-layered phononic crystals with a zero frequency contrast and a non-zero impedance contrast~\cite{nouh2024role}.}
\centering
\begin{tabular}{l l l l}
\hline
Bandgap type & Lower limit &  Upper limit & Width \\
\hline
Wavenumber & $q_- = 2(n-1)\pi + \cos^{-1}(2\nu^2-1)$ &  $q_+ = 2n\pi - \cos^{-1}(2\nu^2-1)$ & $\Delta q = 2\left( \pi-\cos^{-1}\left(2 \nu^2 - 1 \right) \right)$ \\
\hline
Frequency & $\Omega_- = (n-1)\pi + \frac{1}{2}\cos^{-1} \left(2\beta^2-1 \right)$ & $\Omega_+ = n\pi - \frac{1}{2}\cos^{-1} \left(2\beta^2-1 \right)$ & $\Delta \Omega = \pi - \cos^{-1} \left(2\beta^2-1 \right)$ \\
\hline
\end{tabular}
\label{table:functions}
\end{table*}

\section*{Similarities between reversed-sign Willis couplings and bi-layered periodicity}
In regard of the bandgap opening mechanism, a resemblance between reversed-sign periodicity in Willis coupling and phononic crystals of two layers (1 and 2) with special modulation parameters is noticed.~It has been recently shown that identical \textit{frequency} bandgaps in phononic crystals open with perfect periodicity along the \textit{frequency} axis by satisfying two conditions~\cite{nouh2024role}:
\begin{itemize}
    \item A zero frequency contrast, achieved via setting $\ell_1 c_2 = \ell_2 c_1$, and
    \item A non-zero characteristic impedance contrast $\beta = (z_1 - z_2)/(z_1 + z_2)$, where $z_{1,2} = A_{1,2}\sqrt{E_{1,2}\rho_{1,2}}$ are the characteristic impedance of the constitutive layers\footnote{Note that all variables hold identical meaning to what has been defined in the article, and the subscripts 1 and 2 are used to distinguish the first and second layer of the bi-layered phononic crystal.}.
\end{itemize}

For easier comparisons, Table~\ref{table:functions} is compiled to concisely summarize the similarities between the alternating Willis-coupling rod design and the bi-layered phononic crystal in terms of bandgap limits and width.~As inferred from the table, the functions governing the \textit{frequency-bandgap} limits exhibit nearly identical resemblance to their \textit{wavenumber-bandgap} counterpart, which extrapolates to the functions for the bandgaps' width.~Apart from a 2 multiplier for the wavenumber bandgap equations, the two equation sets are identical as if the parameters $\nu$ and $q$ are swapped with $\beta$ and $\Omega$, respectively.~As such, one may deduce that the role of impedance contrast in opening a \textit{frequency} bandgap is equivalent to the role of the normalized modulation speed $\nu$ in opening a \textit{wavenumber} bandgap, given reversed-sign Willis coupling.

%A method for creating wavenumber bandgaps using periodic changes in the signs of Willis coupling is introduced.

\section*{Concluding remarks}
The introduction of Willis coupling in elastic media traditionally yields non-reciprocal band structure due the presence of momentum bias.~Peculiarly, reversing Willis-coupling signs within a unit cell, \textit{ceteris paribus}, is analytically proven to produce reciprocal band structures with wavenumber bandgaps, and analogies to a special class of bi-layered phononic crystals are noted.~Future research directions of the proposed methodology could be focused on physical realizations of periodic Willis coupling.~For instance, while physical motion of an elastic rod gives rise to Willis coupling, synthesizing a feedback controller (e.g., Refs.\cite{pechac2021non,Rosa2020DynamicsInteractions,ALLI2000625}) could be an alternative route for realizing Willis coupling, especially if a periodic variation in Willis coupling is sought. 

\noindent\rule{3.5in}{0.5pt}

% \section*{References}
\footnotesize
\printbibliography[heading=none]
\end{multicols}
\break
\end{document}